# Physics graduate teaching assistants' beliefs about a grading rubric: Lessons learned

Edit Yerushalmi[1], Ryan Sayer[2], Emily Marshman[3], Charles Henderson[4] and Chandralekha Singh[3]

[1]*Department of Science Teaching, Weizmann Institute of Science, 234 Herzl St., Rehovot, Israel 7610001*
[2]*Department of Physics, Bemidji State University, Bemidji, MN, 56601, USA*
[3]*Department of Physics and Astronomy, University of Pittsburgh, Pittsburgh, PA, 15260, USA*
[4]*Department of Physics, Western Michigan University, 1903 W. Michigan Ave., Kalamazoo, MI, 49008*

**Abstract**: Explication and reflection on expert vs. novice considerations within the problem-solving process characterize a cognitive apprenticeship approach for the development of expert-like problem solving practices. In the context of grading, a cognitive apprenticeship approach requires that instructors place the burden of proof on students, namely, that they require explanations of reasoning and explication of problem-solving processes. However, prior research on instructors' considerations when grading revealed their reluctance to use such a grading approach, motivated by a perception of teaching that places the burden of proof on the instructor. This study focuses on physics graduate teaching assistants (TAs) who play a central role in grading. A short professional development activity was designed that involved eliciting TAs' perceptions regarding grading, presenting a cognitive apprenticeship-inspired grading rubric, followed by a discussion of the dilemma between placing the burden of proof on the instructor vs. the student. In this context, we examined TAs' grading considerations and approaches towards grading and the effect of the short professional development grading activity on them.

## I. INTRODUCTION

Goals for many physics courses include helping students develop expert-like problem-solving (PS) practices [1]. A cognitive apprenticeship (CA) approach to learning physics and developing expert-like PS practices involves modeling expert-like approaches to PS, and coaching students and providing them scaffolding support to explicate their PS processes. One way to coach students to develop expert-like PS practices can be via grading practices that weigh the explication and justification of steps in the PS process.

Physics graduate teaching assistants (TAs), especially at large universities, often grade student work. Thus, TAs' beliefs about grading can communicate course goals and expectations to students [2]. Previous research on instructors' grading approaches revealed their tendency to infer student's understanding based on very little evidence and reluctance to grade on the explication of the PS process [3]. Specifically, instructors did not put the burden of proof for explicating the PS process on the students. It is reasonable to expect that TAs, being educated in an environment shaped by instructors, will hold similar attitudes towards grading. We incorporated a unit within a TA professional development (PD) course intended to encourage TAs to have a CA approach toward grading. Following educators' recommendations [4] to provide opportunities to examine collaboratively one's own practice and beliefs, we designed a unit in which TAs graded introductory physics student solutions with and without a CA-inspired grading rubric and engaged in group and class discussions about these experiences. In this context we studied: 1. How do TAs apply a CA-inspired rubric when grading introductory student solutions? 2. What are TAs' stated pros and cons of using a CA-inspired rubric? 3. How do TAs' grading approaches change after the PD unit and one semester of teaching experience?

## II. METHODOLOGY

The study focused on 15 first-year physics graduate TAs participating in a mandatory, semester-long PD course at a research university in the U.S. The majority of the TAs were teaching introductory (intro) physics recitations for the first time. A few were also assigned to facilitate a laboratory section or grade students' work in various physics courses. The PD course met for 2 hours each week and was meant to prepare the TAs for their teaching responsibilities. The TAs were generally asked to do one hour of homework each week pertaining to teaching that was graded for completeness. During class meetings, TAs usually discussed their homework assignment from the previous week in small groups. The instructor then facilitated a class discussion where groups shared their ideas.

One sequence of homework and in-class activities in the PD course involved grading. At the beginning of the semester, TAs were given an intro physics problem and 5 Student Solutions to a core problem (see Fig. 1) drawn from the aforementioned research on instructors' grading [3]. In this study, we focus on two student solutions to the core problem: Student Solution D (SSD) and Student Solution E (SSE) (see Fig. 2). The boxed notes in SSD indicate where explicit mistakes were made. The solutions were designed to reflect common components found in an analysis of several hundred student solutions to the problem as well as the differences between expert and novice PS from the research literature (e.g. qualitative reasoning, planning, evaluation of final answer, etc.) [1]. Both SSD and SSE contain the correct answer. However, the elaborated solution SSD explicates the PS process while the brief solution SSE does not. The TAs were asked to individually grade the student solutions as quizzes out of ten points, list solution features, and justify the

You are whirling a stone tied to the end of a string around in a vertical circle having a radius of 65 cm. You wish to whirl the stone fast enough so that when it is released at the point where the stone is moving directly upward it will rise to a maximum of 23 m above the lowest point in the circle. In order to do this, what force will you have to exert on the string when the stone passes through its lowest point one-quarter turn before its release? Assume that by the time you have gotten the stone going and it makes its final turn around in the circle, you are holding the end of the string at a fixed position. Assume also that air resistance can be neglected. The stone weighs 18 N.

**FIG 1**. The core problem.

**FIG 2**. Student Solutions D (SSD) and E (SSE).

weight they assigned to the solution features to arrive at a final score. TAs were told to assume that they were the instructors of the course and could use any grading method.

After the TAs graded SSD and SSE individually, they discussed their grading approaches in small groups in the next class of the PD course. In a whole class discussion, the TAs shared their grading criteria for grading SSD and SSE. TAs' stated grading criteria generally included listing initial information, drawing a diagram, proof of understanding, errors in physics reasoning, intermediate steps, and correct units. The instructor of the PD course highlighted TAs' stated grading criteria that aligned with a CA approach to grading (that promoted expert-like PS practices) and the disadvantages of grading which focused essentially on correctness. TAs were then given a CA-inspired rubric (see Table I) and asked to discuss the pros and cons of it. The instructor highlighted how the categories of the rubric incorporated the grading criteria mentioned in the whole class discussion (e.g., "list" and "diagrams" as part of initial problem description, "proof of understanding" as explication and justification of principles) and the dilemma between placing the burden of proof on the instructor vs. the student.

**TABLE I**. Rubric used to grade SSD and SSE, including percentages of TAs who selected each category of the rubric and the "experts'" (exp) application of the rubric.

| Sample Grading Rubric | | SSD | | SSE | |
|---|---|---|---|---|---|
| | | % | exp | % | Exp |
| **Problem Description** (2 points) | Comprehensive diagram (+1) | 67 | | 0 | |
| | Diagram is partial (+0.5) | 33 | X | 0 | |
| | Diagram is not present (+0) | 0 | | 100 | X |
| | List is comprehensive (+1) | 20 | | 0 | |
| | List is partial (+0.5) | 67 | X | 0 | |
| | List is not present (+0) | 13 | | 100 | X |
| **Explication** Invoking and justifying principles (2.5 point) | Useful principles are invoked (+0.75 each) | 93 | X | 73 | |
| | Principles that are NOT useful are invoked (+0) | 0 | | 13 | |
| | Principles useful to solve problem are justified (+0.5 each) | 100 | X | 60 | |
| | Principles that are NOT useful are justified (+0.25) | 0 | | 20 | |
| **Conceptual Understanding** (3 points) | Principles applied adequately (+1.5 each) | 33 | | 40 | |
| | Principles applied are partially correct (w/ sign errors, missing terms, etc.) (+0.75 each) | 73 | X | 47 | X |
| **Math Procedures** (1 point) | Algebraic procedures applied adequately | 93 | | 100 | X |
| **Problem Evaluation** (1.5 points) | Evidence of an *attempt to check the reasonability of the solution* (+2, extra credit for checking) | 27 | | 0 | |
| | Answer is *reasonable* and there is *no* evidence of a reasonability check (+1.5) | 73 | X | 87 | X |
| | Answer is *unreasonable* and *no* acknowledgement has been made by student (+0 points) | 0 | | 13 | |
| **Total score** | | - | - | 6.5 | - | 4.0 |

The grading rubric was developed collaboratively by the authors and iterated many times in order to ensure that the content and wording was appropriate. It values expert PS features from the research literature (e.g. problem description, evaluation of final answer, etc.) [1], and was designed to be sufficiently general so that it could be applied to a variety of physics problems. It divides the grading into five separate categories: problem description, explication and justification, conceptual understanding, mathematical procedures, and problem evaluation [5]. The investigators themselves used the rubric to grade SSD and SSE and inter-rater reliability was better than 90%. The scores the investigators agreed upon when using the rubric to grade SSD and SSE are called "expert" scores. The expert score is higher for SSD, indicating the preference for description and explication over correctness. The TAs were asked to grade SSD and SSE again using the CA-inspired rubric. They then reflected on their grading using a rubric by summarizing what they considered pros and cons of using such a rubric to grade student solutions. To investigate whether TAs' grading practices changed after the PD course and a semester of experience as a TA, the TAs were given a homework assignment at the end of semester which asked them to grade SSD and SSE again. They were not given a rubric at this stage, but they were asked to list features of SSD and SSE and explain how they weighed the different solution features in grading. The different grading worksheets were previously developed and content validated by three of the authors in collaboration with peer physics education researchers for use with TAs/instructors [3,6]. TAs' written responses were comprehensive and they actively participated in the in-class discussions, indicating that they took all the grading activities seriously.

After an initial analysis of the collected data, seven of the TAs in the study volunteered to be interviewed to provide further clarification of their grading beliefs and to clarify their written responses on the worksheets. The interview was semi-structured, including pre-determined questions that focused on whether the grading activities carried out in the TA PD course impacted TAs' beliefs about their grading in some manner not captured in their written responses as well possible mismatches between their grading practices using a rubric vs. their actual grading practices in courses in which they graded student work. The interview also included additional follow-up questions.

## III. FINDINGS

Table II shows the average scores assigned for SSD and SSE before the rubric was introduced (pre), the average scores assigned by TAs when using the rubric (rubric), the scores assigned by "experts" using the rubric (expert), and the average scores at the end of the semester after the TAs had completed the grading activities (post), with standard deviations for each average score and the $p$-values for comparison between the means of the pre and post scores. At the beginning of the semester without using a rubric, TAs graded the elaborated solution SSD slightly higher than the brief solution SSE. In addition, Table II shows that TAs' average score on the elaborated solution SSD when using a rubric was approximately the same as their average score on SSD in the pre grading activity. TAs' average score on the brief solution SSE was approximately 1 point lower when using the rubric as compared to the pre grading activity.

However, TAs' scores on SSD and SSE when using a rubric were not in agreement with "expert" graders using a rubric (see Table II). To investigate why the TAs' scores were different from the "experts'" score, we analyzed how the TAs applied the rubric categories when grading SSD and SSE. Table I shows the percentage of TAs who selected each category of the rubric when grading SSD and SSE. Compared to "expert" graders, TAs were generally in agreement with "experts" when grading the categories of problem description, conceptual understanding, mathematical procedures, and problem evaluation. However, TAs were not in agreement with "experts" when grading the brief solution SSE in the explication category. For example, over half of the TAs selected that "useful principles are invoked" and "useful principles are justified" when grading the brief solution SSE, even though there was no evidence of explicit invoking or justifying of physics in the brief solution SSE.

**TABLE II**. Average (Avg.) scores and standard deviations (St. Dev.) assigned to the elaborated (elab.) solution SSD and brief solution SSE before using the rubric (Pre), when using the rubric to grade (Rubric), scores assigned by experts using the rubric, and at the end of the semester (Post), with $p$-values ($p$) for comparison between pre and post scores.

|  |  | Pre | Rubric | Expert | Post | $p$ |
|---|---|---|---|---|---|---|
| SSD (elab.) | Avg. | 7.93 | 7.98 | 6.50 | 8.16 | 0.6 |
|  | St. Dev. | 1.24 | 0.70 |  | 1.60 |  |
| SSE (brief) | Avg. | 7.07 | 6.07 | 4.00 | 7.65 | 0.6 |
|  | St. Dev. | 2.71 | 1.68 |  | 3.10 |  |

To investigate why TAs applied the rubric differently than "expert" graders in the "explication" category of the rubric, we examined TAs' stated pros and cons of the rubric that they listed on the grading worksheet. We read the TAs' written responses to determine whether they followed any trends. Based upon these trends, categories were created to describe the most common types of responses. Two researchers separately coded the responses according to the chosen categories and then compared their individual coding and discussed any discrepancies until an agreement of greater than 90% was reached.

Regarding the pros of using a rubric to grade, many TAs mentioned that the rubric provides fairness/consistency and makes it easier to grade students' solutions. Over 70% of the TAs noted that the CA-inspired rubric encourages students to use good PS practices and over 30% of the TAs said that it allows the instructor to identify specific student difficulties. On the other hand, over half of the TAs mentioned that one con of the rubric was that it was too constraining – it did not allow for enough flexibility when assigning scores (e.g., to give partial credit in certain cases) and did not allow them to assign points the way they would like to. The TAs who mentioned this con were usually uncomfortable in taking off points if the final answer was correct. For example, one TA stated: "The process is one factor, but it's not really that important… I think in most practical cases, the correct answer should be more important than (the process)… that people think (may be important)." In particular, the TAs often felt that they should have the freedom to grade the intro student solution in a manner they see appropriate rather than being tied to a rubric.

Several TAs mentioned that they wanted to be able to give a high score to a student whose final answer was correct even if the student did not explicate their PS approach. One TA stated, "the answers are not like filling in forms. They're much more interwoven and complicated than that. You cannot really say, 'okay, here we have this, so one point to that.' That's not true in the real case. So I just read it (the rubric) and got some idea out of it, but didn't really follow every instruction." Several TAs mentioned in interviews that they would grade based upon their intuition instead of using a rubric if they could infer student understanding from looking at a student's solution. For example, one TA stated: "When students take a quiz, I know that he's not cheating so he knows the answer, but maybe he's stressed or trying to do it really fast, so he did part of it in his mind. I'm sure that he did the right thing for the quiz so I gave him the full grade for the quiz." This TA was willing to give students the benefit of the doubt and infer student understanding because of the time limitation and stress in a quiz context. Another TA mentioned that he identifies with students who write brief solutions, stating: "In my past I've usually answered questions in that form [of a brief solution like SSE], so I guess I can understand what students are trying to say when they write things like that." This TA was among those who gave SSE credit for justifying the use of invoked physics principles when grading with the rubric even though there was no evidence of justification in SSE. This TA and several others noted that when they themselves were students they often submitted brief solutions and were not penalized for not explicating the PS process. TAs with these types of responses typically did not put the burden of proof for explicating the PS on the student.

Other TAs who felt that the rubric was not flexible often did not recognize that grading rubrics that weigh the PS process appropriately can serve as a formative assessment tool for both students and instructors [2]. For example, one TA also claimed that "it is up to the students to get something out of their solution, and student learning should not depend upon the TAs' grading practices." He stated that assigning points to features such as diagrams and lists of unknown variables was merely "sugar coating" the students' scores, i.e., assigning points that inflated student scores and simply helped the students get a better grade but did not help them learn physics. He felt that his students should be like him and attempt to figure out why they performed poorly on their own. He did not agree that grading practices that place the burden of proof on students for explicating the PS process can help students learn physics and develop good PS approaches. Individual interviews and the dilemmas that came up in the discussion in the PD class about the pros and cons of the rubric suggest that this type of feeling was common among TAs.

Table II indicates that TAs' average scores on the solutions SSD and SSE did not change significantly between the beginning (pre) and the end of the semester (post) after working through the grading activities and having one semester of TA experience. It appears that many of the TAs were still willing to infer student understanding in SSE even when there was little explicit evidence of it and did not use a rubric that weighed explanation of reasoning and explication of the PS process suitably.

## IV. SUMMARY

At the beginning of the semester, TAs graded a solution explicating reasoning (SSD) similar to a brief solution (SSE). When TAs were asked to use a CA-inspired rubric that favors explication of problem description and reasoning to grade SSD and SSE, they did not use the rubric as intended when grading the solution in which the final answer was correct but the PS process was not explicated. They were willing to infer understanding even when there was no evidence of it and gave points for explication of reasoning to the brief solution SSE, even though such explication did not take place. After the grading activities in the TA PD course and one semester of experience as a TA, many TAs still did not place the burden of proof on students when grading. Despite class discussions and giving TAs an opportunity to collaborate with other TAs to reflect on their grading approaches, TAs' grading approach did not change significantly by the end of the semester and after one semester of teaching experience.

This study suggests that TAs have a strongly held views about grading that are not aligned with a CA approach for the development of expert-like problem solving. The interviews and class discussions suggest that this approach is anchored in TAs' former experiences: Most TAs, who had not shown the process of arriving at a final answer in their own solutions in the past, had generally managed to get full scores if their answers were correct. Many TAs noted that, throughout their education, their solutions were graded based upon correctness only, and there was often an expectation that if the final answer is correct the student must know how to solve the problem correctly. Other TAs expected that their students were like them and should attempt to figure out why they had performed poorly (if they did so), regardless of how they were graded. They did not think that it was their job to help students develop effective PS approaches, such as explication of the PS process, in order to help students learn.